\documentclass[
]{ceurart}

\usepackage{listings}

\begin{document}

\copyrightyear{2023}
\copyrightclause{Copyright for this paper by its authors.
  Use permitted under Creative Commons License Attribution 4.0
  International (CC BY 4.0).}

\conference{
  C\&ESAR'23: %
  Computer \& Electronics Security Application Rendezvous, %
  Nov. 21-22, 2023, Rennes, France
}

\title{Zero Trust in the Context of IoT: Industrial Literature Review, Trends, and Challenges}

\author[1]{Laurent Bobelin}[%
orcid=0000-0002-3268-4203,
email=laurent.bobelin@insa-cvl.fr,
]
\address[1]{INSA Centre Val de Loire, 88 boulevard Lahitolle 18000 Bourges, France}

\begin{abstract}
 The Zero-trust (ZT) model is an increasingly popular model that relies on the idea that no trust should be granted to any entity (network, persons, devices) by default. ZT model is gaining attention from both research and practice, with various levels of adequation between research developed and real-life applications. NIST provided a standard to fulfill requirements of ZT architecture of network core but many practical aspects remain unspecified, some of them requiring solving first research challenges in order to be implemented efficiently. An example of such an unspecified field is the integration of IoT/Smart Peripheral Devices (SPD). Various reasons explain this gap: specificities of such resources (possibly lower energy/computation power), their lifecycle, and their use, strongly depending on the use of the whole platform IoT devices are part of.

Moreover, additional difficulty to have a good understanding is induced by the fact that both Zero Trust and IoT are identified as promising trends in cybersecurity: many vendors/researchers tag their solutions as IoT integration into the ZT model, with little to no effective compliance to ZT model or standard. Industry is providing many practice-oriented literature, that has to be compared to academic work and standards, in order to consolidate the current state of knowledge and solutions offered to realize this integration. In this paper, we conduct a literature review of non-academic publications, in order to consolidate current knowledge, trends, and future challenges for the industrial integration of IoT devices in ZT architecture.

\end{abstract}

\maketitle

\section{Introduction}

Distributed systems security such as an enterprise or institutional network is usually based on network segmentation: the whole network is divided into different network segments where devices and users accessing it are supposed to have the same privileges. This model, sometimes referred to as the fortress model, has been efficient for securing organizations for decades: using firewalls, it provides efficient means to filter traffic and discard malicious traffic. 

Two major shifts in the way we use distributed systems revealed the drawbacks of this security model: remote access to resources (for telework for example) and the externalization of services through the use of externally provided resources (such as cloud resources).
Indeed within a segment, anyone is implicitly granted the same trust. It means that any compromised resource within a segment is granted full access to this segment and so it can use any possible means to compromise other resources within the same segment. Remote access for telework then opens doors into privileged segments; on the other hand, hosting services using cloud resources gives the possibility for a compromised (malicious or honest but curious) provider to attack the service network segment. 

To enforce security in all those systems, Zero Trust Architecture is the solution progressively adopted by systems designer: it relies on the motto that no trust is implicitly given to any entity within the network (either devices, users, or services). Communication flows are allowed only if an entity, named Policy Decision Point (PDP), agrees to let this flow occur. This entity evaluates the confidence it can have in the different entities involved in a flow based on the knowledge it has about the user, device, and service involved. 

Zero Trust is not a standard tool, but a design motto:  the term is then used by vendors to qualify solutions that respect more or less the recommendations to implement Zero Trust. However NIST standard \cite{NIST} defines the building blocks for a static network, without a federation of identities, and relevant components to deploy. At the core of such an architecture, Trust Engine, Policy Engine, and Policy Enforcement Point, are respectively responsible for trust evaluation, deciding if a flow is authorized or not, and determining how to enforce the decisions. Decisions depend on various information sources and rely on modeling three key entities involved in the communication flow: (1)  the user at the origin of communication, (2) the device from where the communication is originating, and (3) the targeted service. 

Machine to Machine communication breaks the assumption that the user can be identified as a source of vulnerability of the system. Moreover, modern distributed systems also include a wide variety of devices: IoT sensors and actuators, drones, autonomous vehicles (UAV), smartphones, and edge/cloud servers to name a few.  Such systems dynamically evolve: moving devices may change location, and the infrastructure handles continuous flow routing between the moving device and the other entities that are part of the system. Well-known solutions to deal with mobility include 4G MME and 5G AMF, which both have proved their efficiency. But the security provided by those standards is not designed to fit in a Zero Trust architecture, as it relies on an implicit trust in the infrastructure. CISA and NSA provided jointly security guidance based on Zero Trust for the 5G cloud infrastructure\cite{CISA}, leaving the end devices case apart. When dealing with systems that include directly interacting with end-user devices, which may damage the other devices through malicious behavior, it may not be sufficient. This safety concern occurs in various systems, ranging from agricultural systems containing IoT actuators, drones, and autonomous tractors, to military systems. 

Vendors provide solutions to the main challenges induced by IoT-based systems, userless devices, short-lived low resources devices, and black-box devices. GAFAM such as Microsoft Azure \cite{MicrosoftWhitePaper}, along with other major companies such as Zscaler \cite{zscalerwebsite} provides their own solutions, sometimes with little description of the actual enforcement provided, as well as to the respect of the ZT standard. On the other hand, just a few academic works are actually dedicated to overcoming the challenges of applying Zero Trust to systems integrating SPD. This paper aims to provide an industrial literature review of ongoing or recent work in this field. 

The rest of this paper is organized as follows: first, section \ref{S:context} gives an overview of Zero Trust, and IoT specificities that strongly impact the implementation of ZT in such a context. Then in section \ref{s:solutions}, we give the challenges associated to such an implementation, and main trends of how vendors take into account those challenqes. In section \ref{S:industrial}, we give industrial actual solutions to overcome the listed challenges, and conclude in section \ref{S:conclusion}. 


\section{Context}
\label{S:context}
\subsection{Zero Trust}
Zero Trust is a security model relying on the idea that perimeter-based security is inefficient when the so-called perimeter is breached; as nowadays attack campaigns are more and more common, it is likely that a user will compromise at least one of the resources enclosed within this perimeter, and by doing so, will compromise the whole system. It is then mandatory to never grant trust to other resources by default, which is the case in perimeter-based defense. Zero Trust model (ZT, also coined as Zero Trust Network  (ZTN) or Zero Trust Network Architecture (ZTNA) or Zero Trust Architecture (ZTA)), relies on this simple motto "\textit{Never Trust, Always Verify}". 

Since the seminal Google project BeyondCorp \cite{beyondcorp}, the concept has been refined and nowadays can be considered as mature for the industry. NIST provided a recommended architecture \cite{NIST}, and introductory as well as advanced literature targeting administrators \cite{ZTNA} \cite{ZTD} can easily be found. Most major actors in cybersecurity have developed their own solutions. We just briefly introduce the main concepts used later in this paper. Readers may refer to the NIST standard or to the books cited above for a more in-depth vision of ZT.

The ZT model define guidelines to follow \cite{ZTNA}: 
\begin{itemize}
    \item All network flows MUST be authenticated before being processed.
    \item All network flows SHOULD be encrypted before being transmitted.
    \item Authentication and encryption MUST be performed by the endpoints in the
network.
\item All network flows MUST be enumerated so that access can be enforced by the
system.
\item The strongest authentication and encryption suites SHOULD be used within the
network.
\item Authentication SHOULD NOT rely on public PKI providers. Private PKI systems should be used instead.
\item Devices SHOULD be regularly scanned, patched, and rotated.
\end{itemize}

\begin{figure}[ht]
    \begin{center}
       \includegraphics[width=0.85\textwidth]{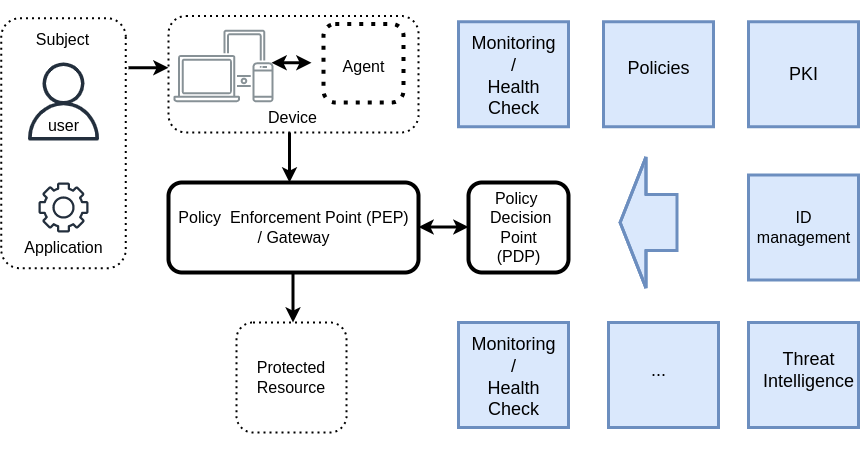}
    \end{center}
\caption{Zero Trust Architecture}
\label{architecture}
\end{figure}

In order to comply to these guidelines, the NIST recommended the logical architecture depicted in figure \ref{architecture}.

The standard identifies the issuer of the request as a \textit{subject} that can be either a (human) user or an application/service. This subject uses a \textit{device} to issue the request. This device may or may not be hosting an \textit{agent}, part of the ZT architecture, that will secure the asset and information provided by this device. The request of access targets a \textit{protected resource} (that may be thought of as data or service). 

\textit{Policy Enforcement Point (PEP)} is often implemented as a gateway: it enforces decisions about whether or not to grant trust to a flow by the \textit{Policy Decision Point (PDP)}. The separation between PEP and PDP relies on the separation between \textit{control plane} (that makes decisions on how to handle the traffic) and \textit{data plane}, widely used in Software Defined Networking (SDN). PEP belongs to the data plane while PDP relies on the control plane. The PDP decision-making is done using as many data sources as possible, in order to make the wisest decision: information about the system state, the users, policies deployed, threat intelligence, etc. 

PDP itself in NIST standard includes the Policy Engine (PE) component, which is the decision-making component, and the Policy Administration (PA) component, which is responsible for coordinating the actions of the PEP so as to reflect the decisions of the PE. Some authors and companies (see for instance \cite{ZTNA}) adds a Trust Engine component (TE). TE is responsible for running a Trust Algorithm (TA), interacting with the different data sources to evaluate risk. PE in this case makes its decision based on the risk evaluation returned by TE and the policies applying to the system. It is then not responsible for evaluating the risk per se. 
Google ZT solution BeyondCorp has pioneered the use of TE: it helps to maintain a lower complexity of the system policy, by discarding edge cases and unknown/unaddressed cases.



\subsection{Specificities of Smart Peripheral Devices With Major Impact on ZT}


Application of Zero Trust to SPD platforms is not straightforward, as it is thought first with the vision of an enterprise network; we here give a review of reasons why IoT use induces challenges for a ZT environment: which component are impacted by these features, how do they impact the different components of the ZT architecture. 


\subsubsection{Userless Devices}
Many SPD  is userless. Implementations of Zero Trust most of the time rely on strong authentication of the user using multi-factor authentication, thus ensuring that the user has access to, at least, a true device, and not a virtual clone. The combination of the trusts granted in both device and user gives a strong authentication, in the sense both user and device are ensuring a mutual authentication: user is guaranteeing that the device used looks genuine, and reciprocally the device guarantees possibly an authentication of user, that can be multi-factorial and/or biometrical. This mutual authentication adds guarantees to the safety of the flow request coming from these 2 entities. When no human/user is part of the process, it is then mandatory to use other means to ensure at least strong authentication. Strong authentication for these userless devices then may rely only on certificates and TPM, as it is done for server resource \cite{C:CESAR2022:SHRfEtCaHtUi}, but may suffer more frequently in the IoT context cloning or replay attacks if not implemented correctly. Low resource combined with possibly black-box devices increase the risk. A company \cite{corsha} developed authenticators to deploy into TPM to increase the number of different authentication factors by the device, and by doing so, achieve MFA for IoT. However, deploying these solutions may be challenging for low-capability resources.

\textbf{Impact:}   Agent, PA and PEP are components strongly impacted by userless devices, as authentication process may require additional, specific measures (userless MFA for example). But as stated upper, the absence of user makes the authentication lower. Lower means of authentication induce that the trust score of those components may be lower compared to standard ones. This is impacting the Trust Engine, Moreover, trust score calculation methods based on user profiling and the request context are by definition not suited to this case. In case a TE is used, the risk evaluation may be done based on the correlation between observed behavior and the one from a digital twin.

\subsubsection{Low Capability} 
By nature, most of IoT devices have low resources (computation, memory, energy). This lack of resource induces low capability for those resources. Low capability of IoT devices may force to use lightweight encryption, which is opposite to the recommended encryption of ZT. It is also challenging, with low-capability devices, to store certificates, and provide TPM to ensure strong authentication. 

\textbf{Impact:}   It impacts the agent implementation, as resources are scarce, and so means to audit the device may be limited. It also limits the possibilities given to the PA to define strong end-to-end enforcement between the IoT resource and the service, thus increasing the risk for the system to be compromised, and necessitates  specific procedures from the PEP. 

\subsubsection{Brownfield Devices} 
Most devices of IoT are provided with legacy software and components, with little to no support to deploy new software or devices. Such legacy devices are sometimes coined as \textit{brownfield devices}, as opposed to \textit{greenfield devices} that allow to deploy software and/or to rely on TPM.
As strong authentication is mandatory and encryption is recommended in ZT, brownfield devices are complex to integrate in a ZT system. 

\textbf{Impact:}   Brownfield devices often impede the implementation of an Agent to deploy on the devices. If agent implementation/deployment is impossible, solutions usually either isolate resources or make use of other sources of trust enforcement techniques in the component (such as digital twin, agentless scanning and asset discovery) or deploying a gateway component that will be responsible for this device. We will discuss in the next section the pro and cons of those solutions. 

\subsubsection{Short-lived Devices} 
Many type of IoT are by nature supposed to be short-lived. Mechanically, the cost of enrolling, maintaining and deploying an agent on those devices is high compared to more long-lived resources. The cost then may be prohibitive to use ZT in this context. 

\textbf{Impact:}   Short-lived devices are one of the resources that fits well with the ZT vision, than includes certificate, possibly complex enrolment, constant monitoring to setup, and potentially long-term profiling of devices. Along with the induced cost of the handling of the shorter life cycle, several problem then may arise when it comes to profile such devices: the low complexity of behavior may make the digital twin inefficient. The massive turn-over of resource may challenge the scalability of the authentication and enrolment systems. Once again a solution that may be considered is the isolation of the resources from other parts of the systems with agentless scanning /asset discovery, with the pro and cons associated with this solution. 

\subsection{Mobility}
SPD includes cellular phones or other complex devices. 4G and 5G address the mobility problem by using either MME (4G) or AMF/network slicing (5G). Both approaches relies on the idea of a trusted network core/infrastructure, that has been recently questioned by the US ban (and possible ban in Europe) of Chinese 5G equipment manufacturers Huawei and ZTE \cite{euronews}.  

\textbf{Impact:}  
The notion of mobility is not considered most of the time in the ZT solutions, as it is not in the ZT initial scope. Mobility impacts the whole chain of component in ZT, as it necessitates adequate authentication, interoperability between systems to achieve end-to-end security, knowledge sharing about access control, monitoring, trust between systems, and so on. 

\subsection{Heterogeneity}
Heterogeneity is one of the aspect of SPD/IoT systems: many sensors and actuators are specifically designed for a given task, and so the diversity in IoT systems may be important. As it is a very active industry, there are a lot of vendors, that multiply the products and variants of them. Moreover, if resource are scarce in those devices, authentication and security implementation may be specifically implemented for a given product version, thus increasing the heterogeneity. 

\textbf{Impact:}  
Heterogeneity is multiplying the possible versions of agent in case of greenfield device, if no standard TPM is provided. In case of brownfield devices, heterogeneity is multiplying the authentication means and implementations. Apart from the increasing attack surface, it also complexify the quantification of trust one can give in authentication and encryption means provided by a given device. It is therefore complicated to write comprehensive policies for a system including a large variety of devices. This risk may be mitigated by the use of a Trust Engine based on Machine Learning, that may embrace automatically this complexity, at the cost of the learning phase that may leave the system relatively vulnerable, even in the case of non-zero day attacks.  Combined with volatility, heterogeneity of devices may lessen the accuracy of ML/Deep Neural Network techniques deployed on TE.

\section{Challenges and Industrial Solutions Proposed}
\label{s:solutions}
From the previously listed specificities, we derived the challenges listed below. After having briefly identified those challenges, we list solutions families proposed by the industry to respond those challenges. 

\subsection{Challenges}

\begin{itemize}
\item \textbf{Agentless Solutions}
Brownfield IoT breaks the ZT assumption that one can deploy an Agent on the device,  breaking the ZT architecture. The challenge here is to grant trust to a device, without fully being able to strongly verify its health, identity, and compromise. It is therefore mandatory, to include such IoT in a ZT system, either to isolate them or to provide agentless solution providing sufficient trust in those devices. Agentless solutions is particularly challenging for the PDP, as it question the way trust can be modeled. Without agent, one may grant lower trust to device and flows coming and outgoing from those devices. To do so, the ZT policy must reflect this lower trust, either by providing means to express this lower trust, or by the definition of restricted access groups for suspicious devices, for example. The core question about agentless solution in the ZT context is then the integration of agentless knowledge in the ZT trust evaluation process.

\item \textbf{Heterogeneity of Trust in Devices} 
Heterogeneity of devices themselves and their possibly low capabilities, combined with the different authentication and encryption capabilities, and the fact that some devices are userless or agentless, multiplies the trust level one can grant to a device: the different authentication means differs in the trust on can grant to. It therefore both complexifies the task of the Trust Engine, as well as the ZT system policy definition by administrators. A good balance must be found between the definition of particular cases definition in policies, increasing the policy complexity but providing clear explainability of the PDP decisions, and the trust the administrator must grant to TE evaluation, in case of use of an non-explainable ML/DNN-based TE.  
\item \textbf{Mobility}
Mobility is a challenge per se, that has been addressed by 4G/5G systems since their beginning. However, those systems relies on the assumption that the network core can be trusted, and that trust can be granted to the different systems where the end user may roam. This break the assumption that a central entity can monitor and secure the whole system, thus delegating the trust to other (possibly non ZT) systems. It then can be considered as a challenge in the way PDP can model and interact with external systems. 

\item \textbf{Interoperability}
Interoperability is a challenging issue in ZT, as interoperability requires that systems grant trust to others. For example, deploying a network component means to trust its provider \cite{CSGFJMŠLP:CESAR2022:PZAfMSSP}. In the case of IoT, one may want to federate heterogeneous agentless systems and/or solutions provided by an industrial to secure those systems. It can be compared to the problem of trust that the system has to grant to agentless devices, but for a whole, external system. 

\item \textbf{Volatility} 
Heterogeneity combined with short-lived devices, constantly renewed, may lead to a volatility of devices, and then a fast turn over in the IoT device types. this may be an issue for PDP as well as PEP, the former necessitating fast update in policies in PE and procedure in PA to adapt to new device types, and the latter necessitating updates in their capabilities to enforce end-to-end security for those devices.  

\end{itemize}

\subsection{Industrial Solutions Proposed}
To overcome these challenges, different solutions and tools have been implemented by vendors. some of them are in all (real) support offered, such as agentless scanning, asset discovery, or classic IoT defense. We give in the following sections an overview of the different types of solutions and tools proposed by industrials.

\subsubsection{Agentless Scanning and Asset Discovery}
Agentless scanning involves the process of evaluating the security posture of IoT devices without requiring the installation of dedicated software agents on each device. Agentless scanning utilizes network-based techniques to identify, assess, and potentially remediate vulnerabilities, configuration weaknesses, or anomalies present in IoT devices. By leveraging existing network infrastructure and protocols, agentless is a scalable method to continuously monitor the security of IoT devices. This approach reduces the burden of managing and updating agents on numerous devices and can streamline vulnerability assessment and management processes in IoT ecosystems. This approach is particularly suited for the diverse and resource-constrained nature of IoT environments. While not offering the control of an agent-based approach, the approach relies on the maturity of such solution in the industry. The approach is supported by many vendors, combined with asset discovery. 

Asset discovery is the systematic process of identifying and cataloging all devices and resources within an IoT ecosystem. Given the sprawling nature and diversity of IoT deployments, in a ZT context, comprehensive asset discovery is essential for effective security management. This process involves actively scanning networks, probing for active devices, and gathering information about their characteristics, communication protocols, and associated metadata. Automated asset discovery tools play a pivotal role in identifying devices that might have been inadvertently added or forgotten, minimizing security blind spots of the ZT system. Asset discovery may be used as the foundation for establishing a robust ZT system.

\textbf{Integration into IoT and ZT system: } 
As it is an existing solution for IoT security, the question is how these tools are integrated to the ZT system. It is mostly about (1) security enforcement by PEP/PA and (2) taking into account at the PDP/TE level the lower trust one can grant to agentless devices. While many vendors claim to offer the support for PEP/PA - that is, as the devices does not include agent, simply initiating communication flow with the strongest encryption offered by the IoT device - we did not find in industrial literature an explicit statement of lower trust granted to IoT systems just integrated using agentless scanning and asset discovery. This may be either let to the vendor's client to write a specific policy for such devices, or supposedly under the TE responsibility to discard anomalous traffic coming from those devices.  

\subsubsection{Digital Twins}
Digital twins are used to predict behavior, and by doing so, evaluate the potential abnormal behavior of a device. While further analysis may be mandatory to determine the root causes of the abnormal behavior, device twins are sometimes associated with quarantine groups that isolate potentially compromised devices. Some vendors may name it those device twins with different names (device shadows for example for AWS). In some configuration, the digital twin may be used as an information source replacing the agent, but with lower trust in the information provided.   

\textbf{Integration into IoT and ZT system: }
Device twin is a trendy concept, not limiting its application scope to IoT systems.  Integration into an IoT system is then quite easy, as the solution is quite mature. However, the same comment as for agentless scanning/asset discovery holds for digital twin, as we did not find any explicit statement of lower trust granted to those devices compared as other.   

\subsubsection{IoT Devices Isolation}
Some vendors actually leave the IoT devices outside de ZT-administrated zone, thus increasing the complexity of the operation of the system. As the shift from perimeter defense to ZT can be incremental by adding progressively the different network segments \cite{ZTNA}, this may be a solution for non-critical IoT sub-systems \cite{sentinelonewhitepaper}. However, in many cyberphysical systems, IoT is the core critical systems.

\textbf{Integration into IoT and ZT system: } 
This solution offers the simplest integration in both systems. Obviously this solution does not provide the security of an integration of IoT system into ZT. 


\subsubsection{Single Device Gateways}
\begin{figure}[]
    \begin{center}
        \includegraphics[width=0.9\textwidth]{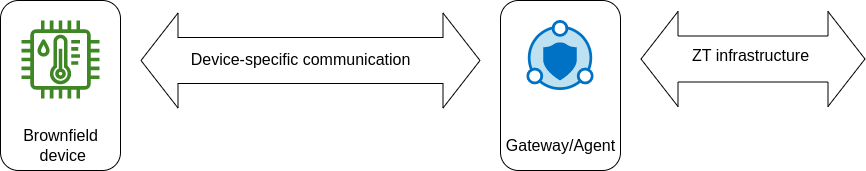}
    \end{center}
\caption{Brownfield device gateway}
\label{l:gateway}
\end{figure}

Gateway is a solution provided by Azure to overcome the brownfield problem. The basic idea is to provide a hardware module responsible for WiFi communication that will either replace or isolate the actual communication module of the brownfield device. The module includes a TPM and an OS, and thus ensures complete compliance of the brownfield device to ZT requirements, as pictured in figure \ref{l:gateway}.

The drawbacks of such an approach are numerous: it increases the attack surface, adds devices to the infrastructure, increases maintenance complexity, and increases energy consumption. Moreover, it may be unfeasible to bypass the WiFi connection of an existing device, thus duplicating the attack surface of a device. 

\textbf{Integration into IoT and ZT system: }
Integrating the device gateway into an existing IoT system may be cumbersome: devices have to be physically modified to include the gateway. However, as it can tunnel the direct communication in between devices, a reconfiguration of the IoT system may not be mandatory. On the ZT systems, the advantage of such an approach is that it offers a seemless integration of brownfield devices. 

\subsubsection{Greenfield-Oriented Software}
Many actors provide SDK in order to implement agents that will be deployed on (greenfield) devices. Most of them imply the support of a TPM and/or a TEE, that will provide support for strong authentication and certificate storage. Azure provides a specific OS (Azure Sphere), others like AWS often recommend the use of some OS (for example FreeRTOS \cite{FreeRTOS}). 

\textbf{Integration into IoT and ZT system: }
In the context of a volatile and heterogeneous IoT platform, the development cost of a dedicated agent may be high . Moreover, deployment may induce physical management of devices. On the ZT system side, it offers the possibility to integrate in the system easily, with the possible complexity induced by the heterogeneity of authentication and encryption methods offered by the devices. 

\subsection{Analysis}
As next section states, most actors, when providing solutions for IoT integration into ZT, use a combination of these solutions. Those approaches mainly address the data plane perspective of the problem, responding to the agentless challenge for the PEP and PA only. 

The challenge of volatility is addressed as non-ZT IoT systems respond it, by agentless scanning and asset discovery. The main question is the integration of the solutions provided, that comes from the IoT  world, into the ZT system: modeling and handling different trust levels to grant to different information sources concerning devices, and including this into a consistent policy.

The TA/TE/PDP problem of the  heterogeneity of trust in devices is the same underlying problem common to interoperability, mobility: the modeling/handling of various level of trust to grant to different device types. It is most of the time omitted from the industrial literature, with the exception of quarantine groups in AWS solution, that allows to restrict access on a per device base, waiting for an administrator to either stop restriction or stop the attack. (Manual) policy definition may also allow the administrator to define such quarantine groups, at the expensive cost of administrator (non error-proof) work. It is unclear from many vendors documentation if this problem is ignored and let under the administrator responsibility, or if TE is supposed to be able to efficiently deal with it.

In the case it is ignored, it means that there is an important work to do to take into account those challenge and enable mobility and interoperability of ZT+IoT systems. In the case this problem is supposed to be solved by TE, it questions its role: TE is supposed to marginally help the policy by discarding and managing blind spots in policies, not to handle a large part of the ZT security, where non-explainable automatic decisions may lead to instability of the system. 

Finally, implementing mobility and interoperability of ZT+IoT system may also require specific handling by the data plane ; interoperability between different ZT systems may require the adaptation to ZT mechanics used for the interoperability of security between different IoT system such as the one specified by the OneM2M framework \cite{ONEM2M}.  
%

\section{Overview of Main Industrial Solutions}
\label{S:industrial}

In this section we provide an analysis of the solutions of the 17 main actors we identified as possible providers of ZT+IoT solutions. After the description of our methodology to identify actors in section \ref{ss:identification}, we provide an overview of the different solution in section \ref{ss:overview}. We finally provide a short description of solutions for actors providing support for ZT+IoT in section \ref{ss:providing}, and list the actors not providing support for ZT+IoT in section \ref{ss:not}.

\subsection{Identification of Main Actors}
\label{ss:identification}
Zero Trust, as well as IoT, are both really active domains, both in terms of marketing and actual use. There is therefore a plethora of offers, with various credibility in terms of reality to their offers and effective compliance to ZT model and support for IoT. We used 4 search criteria to identify industrial actors: (1) Top actors in terms of revenue in cybersecurity (2) Major actors of Cloud-based hosting (3) GAFAM (4) Renowned actors in cybersecurity. 

We used 2023 ranking in terms of revenue to identify the 10 main actors in this domain \cite{top10cyber}, namely Microsoft Corporation, IBM, Cisco Systems, Inc., Oracle Corporation
, Juniper Networks, Synopsys, Palo Alto Networks, McAfee, Fortinet, CyberArk. 

As of 2022 top 10 major cloud provisioners \cite{top10Cloud} includes Amazon Web Services (AWS), Microsoft Azure, Google Cloud Platform (GCP), Alibaba Cloud, Oracle Cloud, IBM Cloud (Kyndryl), Tencent Cloud, OVHcloud, DigitalOcean, and Linode (Akamai). Only the 4 firsts are providers with more than 5\% of market share. Out of GAFAM only Facebook and Apple are not actors of cloud provisioning and are then discarded. 

This let us with the following 17 actors: Microsoft (Azure), IBM Cloud (Kyndryl),  Oracle, Juniper, Synopsys, Palo Alto Networks, McAfee, Fortinet, CyberArk, AWS, Google (GCP, BeyondCorp), Alibaba Cloud, Tencent Cloud, OVHCloud, DigitalOcean, Linode (Akamai). We added to the list Zscaler, SentinelOne, and NetFoundry as renowned actors, based on the internet research we have done with the sentence \textit{Zero trust IoT}. We give here an overview of solutions - when existing - of the actors in this list, and give a brief overview of what we found when the actors did not provide nowadays solutions to integrate IoT in ZT architecture. 

\subsection{Solutions Overview}
\label{ss:overview}
From our list, and despite the intensive communication, we found out that only 5 out of the 17 actors actually provide support for ZT+IoT, if we exclude guidelines and white papers. Those 5 actors are Azure, Palo Alto  Networks, Fortinet, AWS and NetFoundry. 

Any of those 5 actors provide a common core features for IoT: agentless scanning and asset discovery. The level of integration of those standard IoT tools into ZT architecture is unclear in most solution description. It is however explicitly stated in AWS documentation that there is some integration between the agentless scanning and the ZT security. 

Device twin is explicitly stated as a provided feature by all solutions except for Netfoundry. As well, AWS is the only product that mention integration of this feature in the ZT tools provided. 

Greenfield SDK is provided by Microsoft,  AWS and NetFoundry. We did not find such tool for Fortinet, while Palo Alto white paper state explicitly their solutions is agentless.  

Finally, Azure is the only solution providing IoT hardware for both greenfield and brownfield devices. The most complete solution - in the sense it covers most of the use case - seems then to be Azure. However, from the description given by actors, AWS and Palo Alto seems to provide better integration of their ZT and IoT tools, as those actors both provide either hints and/or technical details about their integration.  

In general, while there is an addition of tools to manage either ZT or IoT, there is no emphasis - and sometimes no mention - on how those tools are integrated. It is unclear if those system simply coexists or does form a consistent system - with the exception of NetFoundry. NetFoundry is the only documentation that clearly integrates IoT and ZT tools, by providing means to coordinate greenfield SDK, brownfield agentless solutions, and ZT policies.  

\subsection{Actors Providing Support for IoT in a ZT Context}
\label{ss:providing}
\subsubsection{Azure}
Azure proposes a whole ecosystem, ranging from IoT hardware to Cloud services to provide ZT for IoT. In their white paper \cite{MicrosoftWhitePaper}, the approach recommended is to start from an existing Cloud/Edge ZT-secured infrastructure relying on Azure tools, then evaluate which of your IoT devices belongs to greenfield/brownfield, and then deploy solutions based on their chipsets. 

Tools provided to implement Cloud/Edge Azure-based ZT IoT includes tools to manage your IoT devices, secure your communications and updates, and SIEM/SOAR. To secure the IoT devices, Azure proposes two different hardware: one dedicated to greenfield, called Azure Sphere, which is a hardware and OS relying on a TPM to manage strong trust in the device itself, and Azure Sphere Guardian, which is a hardware module to integrate on top of brownfield devices. The module acts as a gateway to connect the brownfield device to the network and includes both chipsets and OS to act like the greenfield devices.

\subsubsection{Palo Alto Networks}
Palo Alto Networks is a cybersecurity enterprise, working for strategic organisms like DoD. They describe their support for IoT in ZT environment in a white paper \cite{paloaltowhitepaper}. They do so by providing tools to discover and assess risk, enforcing least (network) access, and continuous monitoring. Anomaly detection techniques include device twins. 

\subsubsection{Fortinet}
Fortinet \cite{fortinet} is a cybersecurity enterprise providing various solutions, including ZT \cite{fortinetzt}. Their solution brief claims to support IoT integration into ZT architecture, by providing using FortiNAC \cite{fortinac} for network access control, an agent solution named FortiClient, and usual tools for asset discovery and vulnerability detection.

\subsubsection{AWS}
AWS IoT Core provides means to connect IoT to AWS Cloud, based on MQTT protocol.  AWS approach does not include physical devices to secure IoT. It relies on the assumption that devices will be able to provide strong authentication using x509 certificates or other legacy means. AWS provides a client to interact with the AWS platform. AWS Device Defender is used to realize audit, and one can use device shadows to detect misbehavior from your device and/or synchronize the device state. Based on anomaly detection of behavior detected at AWS IoT Core using rules defined by the administrator, possibly compromised IoT devices can be put in a quarantine group with limited access to other resources.  

\subsubsection{NetFoundry}
NetFoundry \cite{netfoundry} is a company developing both free, open source software as well as legacy one. They aim to provide Zero trust networking to a large set of devices and entities, from servers and services to IoT. Their support of IoT consists in providing SDK for greenfield devices and address the problem of brownfield devices by providing usual agentless scanning \cite{netfoundryiot} and asset discovery. Their SDK relies on OpenZiti, a programmable network overlay and associated edge components for application-embedded, zero-trust networking \cite{openziti}. Their solution differs significantly from other, as it provides a clear integration of IoT and ZT tools into a consistent framework. 

\subsection{Actors not Providing Support for IoT in ZT Context}
\label{ss:not}

\textbf{Beyond Corp} Google designed its ZT tool suite, BeyondCorp \cite{beyondcorp}, in a agentless manner, without the IoT use case in mind. While large SPD will be used seamlessly in BeyondCorp, there is little to no support for IoT. 
BeyondCorp design relies on the portal-based architecture \cite{NIST} derived from ZT \cite{43231}. In BeyondCorp, HTTPS browser-based access to services is assumed; it is then not designed to handle machine-to-machine communications.

\textbf{IBM Cloud} does not seem to offer solutions for ZT-IoT based system, while their research blog and web site do contain statements concerning the mandatory use of ZT in IoT context \cite{ibmblog}, \cite{ibmweb}. Kyndril (IBM spin-off) does provide tools to secure SPD endpoints, based on asset discovery, agentless scanning, vulnerability detection \cite{kyndryl}.

\textbf{Oracle Cloud} is the solution offered by Oracle to achieve Zero Trust in their infrastructure. While providing guidelines on how to implement ZT in their infrastructure, they do not mention IoT in their white paper dedicated to IoT \cite{oraclewhitepaper}.

\textbf{Juniper Network} provides network, data center-oriented ZT solutions \cite{juniper}; according to their website, they only support IoT protection using threat intelligence \cite{juniperthreat}.

\textbf{Synopsys, inc} \cite{synopsys} is a company that focuses on silicon design and verification, silicon intellectual property, and software security. While they do promote the use of the ZT model and do provide secure IoT chips, it seems both activities are independent of each other.

\textbf{McAfee} \cite{mcafee} is a company well-known from the general public. They have announced they have developed their own ZT solution in 2021, and were at that time actually providing some IoT-related defense tools. Since 2021 they have been bought by Symphony Technology Group, and their activities dedicated to enterprise have been split in 2022 between Trellix and Skyhigh Security, the latter providing ZT solutions without mentioning IoT as part of their targeted platform \cite{skyhighwhitepaper}. It seems that their support has stopped. 


\textbf{CyberArk} \cite{cyberark} is a cybersecurity enterprise specializing in identity management. In their white paper about Zero Trust \cite{cyberarkwhitepaper} they do not mention specific support for IoT. 

\textbf{Alibaba Cloud} offers both ZT guidelines \cite{alibabacloudzt} and IoT solutions \cite{alibabacloud}, but the solutions are independent.  

\textbf{OVHCloud} offers ZT support \cite{ovhcloud} but does not mention IoT. 

\textbf{DigitalOcean} \cite{dg} relies on NetFoundry solutions \cite{dgzt} for Zero Trust.

\textbf{Akamai} offers ZT guidelines \cite{akamai}  but do not provide ZT support.  

\textbf{SentinelOne} \cite{sentinelOne} is a company selling Machine Learning based cybersecurity. In their Zero Trust guidelines, they specify that IoT base should not be considered in the ZT system \cite{sentinelonewhitepaper}.  

\textbf{Zscaler} \cite{zscalerwebsite} is a company selling  a Zero Trust solution named Zero Trust Exchange \cite{ZscalerExchanger}. While there are some hints about how their solution works, the architecture is not disclosed. They seem to have the same portal-based as the BeyondCorp solution. The only support we found for IoT is the ability to have remote privileged access to IoT based on roles. 



\section{Conclusion and Future Works}

In this paper, we gave an overview of the main industrial solutions to integrate Iot/SPD devices in ZT-secured systems. This work is actually a snapshot of the offered solutions in early 2023, and as the topic is quite active, solutions may evolve quickly. It appears that by now the support is really varying depending on the actors, ranging from no support to complete one, including the whole hardware and software chain. Most of the time, in the main actors list we considered, no support is offered for ZT for IoT, while most websites mention the importance of applying ZT principles to IoT. 

Future work will be to do an academic literature review on this topic; by now the domain is not intensively covered, but it is expected that the attention on Zero Trust rise drastically in the near future. Based on some preliminary academic paper reviews we have done, there is a significant gap between the actual solutions proposed by the industry and the research produced, both in terms of subjects that are considered challenging and in the actual integration of the work into a whole ecosystem.

\label{S:conclusion}
\acknowledgments
MERIAVINO is part of the ERA-NET Cofund ICT-AGRI-FOOD, with funding provided by national sources [ANR, UEFISCDI, GSRI] and co-funding by the European Union’s Horizon 2020 research and innovation program, Grant Agreement number 862665.
\bibliography{sample-ceur}




\end{document}